\pdfoutput=1
\PassOptionsToPackage{table}{xcolor}
\documentclass[sigconf,screen]{acmart}

\usepackage{multirow}
\usepackage{enumitem}
\usepackage{bigstrut}
\usepackage[tikz]{bclogo}
\usepackage[skins]{tcolorbox}
\usepackage{wrapfig}
 \usepackage[flushleft]{threeparttable}

\usepackage[usestackEOL]{stackengine}
\usepackage{xcolor, graphicx}

\usepackage[final]{listings}
\usepackage{algorithm}
\usepackage[noend]{algpseudocode}

\AtBeginDocument{%
  \providecommand\BibTeX{{%
    \normalfont B\kern-0.5em{\scshape i\kern-0.25em b}\kern-0.8em\TeX}}}

\newcommand{\IT}[1]{{\bf%
DODGE(\ifx*#1$\mathcal{E}$\else#1\fi)}}

\newcommand{\bi}{\begin{itemize}[leftmargin=0.4cm]}
	\newcommand{\ei}{\end{itemize}}
\newcommand{\be}{\begin{enumerate}[leftmargin=0.4cm]}
	\newcommand{\ee}{\end{enumerate}}

\begin{document}

\title{Fairway: A Way to Build Fair ML Software}
 
\author{Joymallya Chakraborty}
\email{jchakra@ncsu.edu}
\affiliation{%
\institution{North Carolina State University}
\city{Raleigh}
\country{USA}}

\author{Suvodeep Majumder}
\email{smajumd3@ncsu.edu}
\affiliation{%
\institution{North Carolina State University}
\city{Raleigh}
\country{USA}}

\author{Zhe Yu}
\email{zyu9@ncsu.edu}
\affiliation{%
\institution{North Carolina State University}
\city{Raleigh}
\country{USA}}

\author{Tim Menzies}
\email{timm@ieee.org}
\affiliation{%
\institution{North Carolina State University}
\city{Raleigh}
\country{USA}}

\begin{abstract}

Machine learning software is increasingly being used to make decisions that affect people's lives. But sometimes, the core part of this software
(the learned model), behaves in a biased manner that gives undue advantages to a specific group of people  (where those groups are determined by  sex, race, etc.). This \textit{``algorithmic discrimination''} in the AI software systems has become a matter of serious concern in the machine learning and software engineering community. There have been works done to find ``algorithmic bias'' or ``ethical bias'' in software system. Once the bias is detected in the AI software system, mitigation of bias is extremely important. In this work, we \textbf{a)} explain how ground truth bias in training data affects  machine learning model fairness and how to find that bias in AI software, \textbf{b)} propose a method \textbf{Fairway} which combines pre-processing and in-processing approach to remove ethical bias from training data and trained model.
Our results show that we can find bias and mitigate bias in a learned model, without  much damaging the predictive performance of that model. We propose that (1)~testing for bias and (2)~bias mitigation should be a routine part of the machine learning software development life cycle. Fairway offers much support for these two purposes.

\end{abstract}


\copyrightyear{2020}
\acmYear{2020}
\setcopyright{acmcopyright}\acmConference[ESEC/FSE '20]{Proceedings of the 28th ACM Joint European Software Engineering Conference and Symposium on the Foundations of Software Engineering}{November 8--13, 2020}{Virtual Event, USA}
\acmBooktitle{Proceedings of the 28th ACM Joint European Software Engineering Conference and Symposium on the Foundations of Software Engineering (ESEC/FSE '20), November 8--13, 2020, Virtual Event, USA}
\acmPrice{15.00}
\acmDOI{10.1145/3368089.3409697}
\acmISBN{978-1-4503-7043-1/20/11}

\maketitle

\section{Introduction}
Software plays an important role in many high-stake applications like finance, hiring, admissions, criminal justice. For example, software generates models that decide whether a patient gets released from hospital or not \cite{Medical_Diagnosis,7473150}.
Also, software helps us to choose what products to buy \cite{Wall_Street}; which  loan applications are approved \cite{forbes};
which citizens get   bail or     sentenced to jail \cite{propublica}. Further,
self-driving cars are run by software which may lead to damage of property or human injury \cite{Self_Driving}.    These all are examples of     software systems where the core part is machine learning model.

One problem with any  machine learning (ML) model is they are all  a form of statistical  discrimination. Consider, for example, the discriminatory nature of  
decision tree learners that deliberately selects attributes
to divide that data into different groups.
Such  discrimination  becomes unacceptable and unethical when it gives certain privileged groups advantages while disadvantaging other   unprivileged groups (e.g. groups divided by age, gender, 
skin color, etc). In such situations, discrimination or bias is not only objectionable, but illegal. 

Much recent  SE researchers  presume that   the construction of fairer, less biased
AI systems is a research problem for software engineers~\cite{Angell:2018:TAT:3236024.3264590,Brun2018SoftwareF}. 
We assert that modern principles for software engineering should encompass principles for building AI/ML software. This paper mainly focuses on improving AI software  to satisfy an important and specific non-functional requirement - \textbf{fairness}. In the age of agile software development, requirements gathering, architectural design, implementation, testing, verification - in any step, bias may get injected into software system. So, test and mitigation is now a primary concern in any SE task that uses AI. 

Many researchers agree  that fairness is a SE problem
worthy of SE research.
For example, entire conference series are now dedicated to this topic: see the ``Fairware'' series\footnote{http://fairware.cs.umass.edu};
the ACM FAT conference  FAT \cite{FAT}   (``FAT'' is short for fairness, accountability, and transparency); and the IEEE ASE EXPLAIN  \cite{EXPLAIN} workshop series.
Nevertheless,   when discussing this work with colleagues, we are still (sometimes)
asked if this problem   {\em can} or {\em should}
 be addressed by 
software engineers. We reply that:
\bi
\item
SE researchers
{\em can} address bias mitigation.
As shown below, technology developed within the SE
community can be applied to reduce  ML bias.
\item As to whether or not this community {\em should}
explore ML bias mitigation, that is no longer up to us. 
When users discover problems with software, it is the job of
the person maintaining that software (i.e.  a software engineer) to fix that problem.
\ei
For all these reasons, this paper explores ML bias mitigation. In the recent software engineering literature, we have found some works to identify bias in machine learning software systems \cite{Angell:2018:TAT:3236024.3264590, Aggarwal:2019:BBF:3338906.3338937}. But there is no prior work done to explain the reason behind the bias and also removing the bias from the software. We see some recent works from ML community to mitigate ML model bias. All of these works trust the ground truth or the original labels of the training data. But any human being or algorithm can make biased decisions and introduce biased labels. For example, white male employees were given higher priority to be selected for company leadership by human evaluators \cite{Fortune}; COMPAS Recidivism algorithm was found biased against black people\cite{propublica}. If these kind of biased data is used for machine learning model training, then trusting the ground truth could introduce unfair decisions in future. So, training data validation, testing model for bias and bias mitigation are equally important. This paper covers all the concerns. The idea of \textit{Fairway} comes from two research directions:

\bi

\item Chen et al. mentioned that a model acquires bias from training data\cite{chen2018classifier}. They bolstered on data collection process and training data sampling. Their work motivated us to find bias in the training data rather than model.  

\item Berk et al. have stated that achieving fairness has a cost \cite{berk2017convex}. Most of the bias mitigation algorithms damage the performance of the prediction model while making it fair. This is called \textit{accuracy-fairness trade-off}. When trading off competing goals, it is useful to apply multiobjective optimization. While doing so one objective is to reduce bias or achieve fairness and another objective is to keep the performance of the model similar. 

\ei

Drawing inspiration from both these works, we propose a new algorithm, ``Fairway'', which is a combination of pre-processing  and in-processing methods. Following the motivation of Chen et al, we evaluate the original labels of the training data and identify biased data points which can eventually make the machine learning model biased. Then following the idea of Berk et al, we apply multiobjective optimization approach to keep the model performance same while making it fair. The combination of these two approaches makes Fairway a handy tool for bias detection and mitigation. Overall, this paper makes the following contributions:
\bi

\item We explain how a machine learning model acquires bias from training data. 

\item We find out the specific data points in training data which cause the bias. Thus, this work includes finding bias in AI software.

\item We are first to combine two bias mitigation approaches - pre-processing (before model training) and in-processing(while model training). This combined method, \textit{Fairway}, performs better than each individual.

\item Our results show that we can achieve fairness without much damaging the performance of the model. 

\item We comment on the shortcomings of broadly used fairness metrics and how to overcome that.

\item We describe how concept of ethical bias depends on various applications and how we can use different fairness definitions in different domains.   

\item Our Fairway replication package is publicly available  on GitHub\footnote{https://github.com/joymallyac/Fairway} and figshare\cite{Chakraborty2020artifact}. This last point is not
so much a research contribution but a systems contribution since it
enables other researchers to repeat/confirm and perhaps even refute/improve
our results.
\ei
 
 The rest of this paper is structured as follows- Section \ref{Background} provides an overview of software fairness and generates the motivation of this work. Two subsections summarize the previous works. Section \ref{Terminology} explains some fairness terminology and metrics. Section~\ref{Datasets} describes the five datasets used in our experiment. Section \ref{Methodology} describes our methodology to make fairer software. Section \ref{Results} shows the results for six research questions. In section \ref{Threats_to_Validity}, we have stated the threats to validity of our work.  Finally Section~\ref{Conclusion} concludes the paper.

\section{Background}
\label{Background}
\subsection{About Software Fairness}
\label{Software_Fairness}

There are  many instances of a machine learning software being biased and generating arguably unfair decisions. Google's sentiment analyzer model is used to determine positive or  negative  sentiment. It gives negative  score  to  some sentences like `I  am  a  Jew',  and  `I  am  homosexual' \cite{Google_Sentiment}. Google's photo tagging software mis-categorizes dark-skinned people as animals \cite{Google_Photo}. Translation engines inject social biases, like, ``She is an engineer, He is a nurse'' translates into Turkish and back into English becomes ``He is an engineer, She is a nurse'' \cite{Caliskan183}. A study was done on YouTube's automatically-generated captions across two genders. It is found that YouTube is more accurate when automatically generating captions for videos with male than female voices \cite{tatman-2017-gender}. A popular facial-recognition software shows error rate of 34.7\% for dark-skinned women and 0.8\% for light-skinned men \cite{Skin_Bias}. Recidivism assessment models that are used  by  the  criminal  justice  system  have been found to be more likely to falsely label black defendants as future criminals at almost twice the rate as white defendants \cite{Machine_Bias}. Amazon scraped automated recruiting tool that showed bias against women \cite{Amazon_Recruit}.











In 2018, Brun et al. first commented that it is now time that software engineers should take these kinds of discrimination as a major concern and put effort to develop fair software \cite{Brun2018SoftwareF}. A software is called fair if it does not provide any undue advantage to any specific group (based on race, sex) or any individual. This paper represents a method \textbf{Fairway} which  specifically tries to detect and mitigate ethical bias in a binary classification model used in many AI software. 

\subsection{Previous Work}
\label{Background And Previous Work}

Bias in Machine Learning models is a well-known topic in ML community. Recently, SE community is also showing interest in this area. Large SE industries have started putting more and more importance on ethical issues of ML model and software. IEEE \cite{IEEEethics}, the  European Union \cite{EU} and Microsoft \cite{MicrosoftEthics} recently published the ethical principles of AI. In all three of them, it is stated that an intelligent system or machine learning software must be fair when it is used in real-life applications. IBM has launched a software toolkit called  AI Fairness 360 \cite{AIF360} which is an extensible open-source library containing techniques developed by the research community to help, detect and mitigate bias in machine learning models throughout the AI application lifecycle. Microsoft has created a research group called FATE \cite{FATE} which stands for Fairness, Accountability, Transparency, and Ethics in AI. Facebook announced they developed a tool called Fairness Flow \cite{Fairness_Flow} that can determine whether a ML algorithm is biased or not. ASE 2019 has organized first International Workshop on Explainable Software \cite{EXPLAIN} where issues of ethical AI were extensively discussed. German et al. have studied different notions of fairness in the context of code reviews\cite{10.1145/3180155.3180217}. In summary, the importance of fairness in software is rising rapidly. So far, the researchers have concentrated on two specific aspects -  

\bi
\item Testing AI software model to \textbf{find ethical bias}
\item Making the model prediction fair by \textbf{removing bias}
\ei 

\subsection{ Finding Ethical Bias } 
Angell et al. \cite{Angell:2018:TAT:3236024.3264590}  commented that software fairness is part of software quality. An unfair software is considered as poor quality software. Tramer and other researchers proposed several ways to measure discrimination \cite{Tramer_2017}.  Galhotra et al. created THEMIS \cite{Galhotra_2017}, a testing-based tool for measuring how much a software discriminates, focusing on causality in discriminatory behavior. THEMIS selects random values from the domain for all the attributes to determine if the system discriminates amongst the individuals. Udeshi et al. have developed AEQUITAS \cite{Udeshi_2018} tool that automatically discovers discriminatory inputs which highlight fairness violation. It generates test cases in two phases. The first phase is to generate test cases by performing random sampling on the input space. The second phase starts by taking every discriminatory input generated in the first phase as input and perturbing it to generate furthermore test cases. Both techniques THEMIS and AEQUITAS aim to generate more discriminatory inputs. The researchers from IBM Research AI India have proposed a new testing method for black-box models \cite{Aggarwal:2019:BBF:3338906.3338937}. They combined dynamic symbolic execution and local explanation to generate test cases for non-interpretable models.

These all are test case generation algorithms that try to find bias in a trained model. We did not use these methods because along with the model, we also wanted to find bias in the training data. We developed our own testing method based on the concept of situation testing\cite{10.5555/3060832.3061001}.

\subsection{ Removing Ethical Bias }
\label{Removing_Bias}
The prior works in this domain can be classified into three groups depending on the approach applied to remove ethical bias. 

\bi
\item \textbf{Pre-processing algorithms}: In this approach, before classification, data is pre-processed in such a way that discrimination or bias is reduced. Kamiran et al. proposed \textit{Reweighing} \cite{Kamiran2012} method that generates weights for the training examples in each (group, label) combination differently to achieve fairness. Calmon et al. proposed an \textit{Optimized pre-processing} method \cite{NIPS2017_6988} which learns a probabilistic transformation that edits the labels and features with individual distortion and group fairness.

\item \textbf{In-processing algorithms}: This is an optimization approach where the dataset is divided into three sets - train, validation and test set. After learning from training data, the model is optimized on the validation set and finally applied on the test set. Zhang et al. proposed \textit{Adversarial debiasing}  \cite{zhang2018mitigating} method which learns a classifier to increase accuracy and simultaneously reduce an adversary's ability to determine the protected attribute from the predictions. This leads to generation of fair classifier because the predictions cannot carry any group discrimination information that the adversary can exploit. Kamishima et al. developed \textit{Prejudice Remover} technique \cite{10.1007/978-3-642-33486-3_3} which adds a discrimination-aware regularization term to the learning objective of the classifier.

\item \textbf{Post-processing algorithms}: This approach is to change the class labels to reduce discrimination after classification. Kamiran et al. proposed \textit{Reject option classification} approach \cite{Kamiran:2018:ERO:3165328.3165686} which gives favorable outcomes to unprivileged groups and  unfavorable outcomes to privileged groups within a confidence band around the decision boundary with the highest uncertainty. \textit{Equalized odds post-processing} is a technique which particularly concentrate on the Equal Opportunity Difference(EOD) metric. Two most cited works in this domain are done by Pleiss et al. \cite{pleiss2017fairness} and Hardt et al \cite{hardt2016equality}.  

\ei

Fairway combines both \textit{Pre-processing} and \textit{In-processing} approach. Further, post-processing is not needed after using Fairway. Changing a misclassified label requires domain knowledge based on the type of application.  That kind of knowledge can be difficult to collect (since it requires access to subject matter experts).
Hence, post-processing is not explored in this paper. 

\section{Fairness Terminology}
\label{Terminology}

In this section some specified terminology from the field of fairness in machine learning are described. This paper is limited to the binary classification models and tabular data(row-column format). Each dataset used has some attribute columns and a class label column. A class label is called \textit{favorable label} if its value corresponds to an outcome that gives an advantage to the
receiver. Examples include - being hired for a job, receiving a loan. \textit{Protected attribute} is an attribute that divides a population into two groups (privileged \& unprivileged) that have difference in terms of benefits received. An example of such attribute could be ``sex'' or ``race''. These attributes are not universal but are specific to the application. \textit{Group fairness} is the goal that based on the protected attribute, privileged and unprivileged groups will be treated similarly. \textit{Individual fairness} is the goal of similar individuals will receive similar outcomes.

\section{Fairness Measures}

Martin argues (and we agree) that  ``bias is a systematic error'' \cite{bias_systemetic}. Our main concern is unwanted bias that puts privileged groups at a systematic advantage and unprivileged groups at a systematic disadvantage. A \textit{fairness metric} is a quantification of unwanted bias in models or training data \cite{IBM}. We used two such fairness metrics in our experiment-

\begin{table}[]
\caption{Combined Confusion Matrix for Privileged(P) and Unprivileged(U) Groups.}
\label{Confusion_Matrix}
\begin{tabular}{|c|c|c|c|c|}
\hline
\rowcolor[HTML]{C0C0C0} 
           & \begin{tabular}[c]{@{}c@{}}Predicted\\ No\end{tabular} & \begin{tabular}[c]{@{}c@{}}Predicted\\ Yes\end{tabular} & \begin{tabular}[c]{@{}c@{}}Predicted\\ No\end{tabular} & \begin{tabular}[c]{@{}c@{}}Predicted\\ Yes\end{tabular} \\ \cline{2-5} 
\rowcolor[HTML]{C0C0C0} 
           & \multicolumn{2}{c|}{\cellcolor[HTML]{C0C0C0}Privileged}                                                          & \multicolumn{2}{c|}{\cellcolor[HTML]{C0C0C0}Unprivileged}                                                        \\ \hline
Actual No  & {$TN_P$}                                                     & \multicolumn{1}{c|}{$FP_P$}                                 & {$TN_U$}                                                     & {$FP_U$}                                                      \\
Actual Yes & {$FN_P$}                                                     & \multicolumn{1}{c|}{$TP_P$}                                 & {$FN_U$}                                                     & {$TP_U$}                                                      \\ \hline
\end{tabular}
\end{table}

\begin{figure}[t]
\centering
\includegraphics[width=\linewidth,height=5cm]{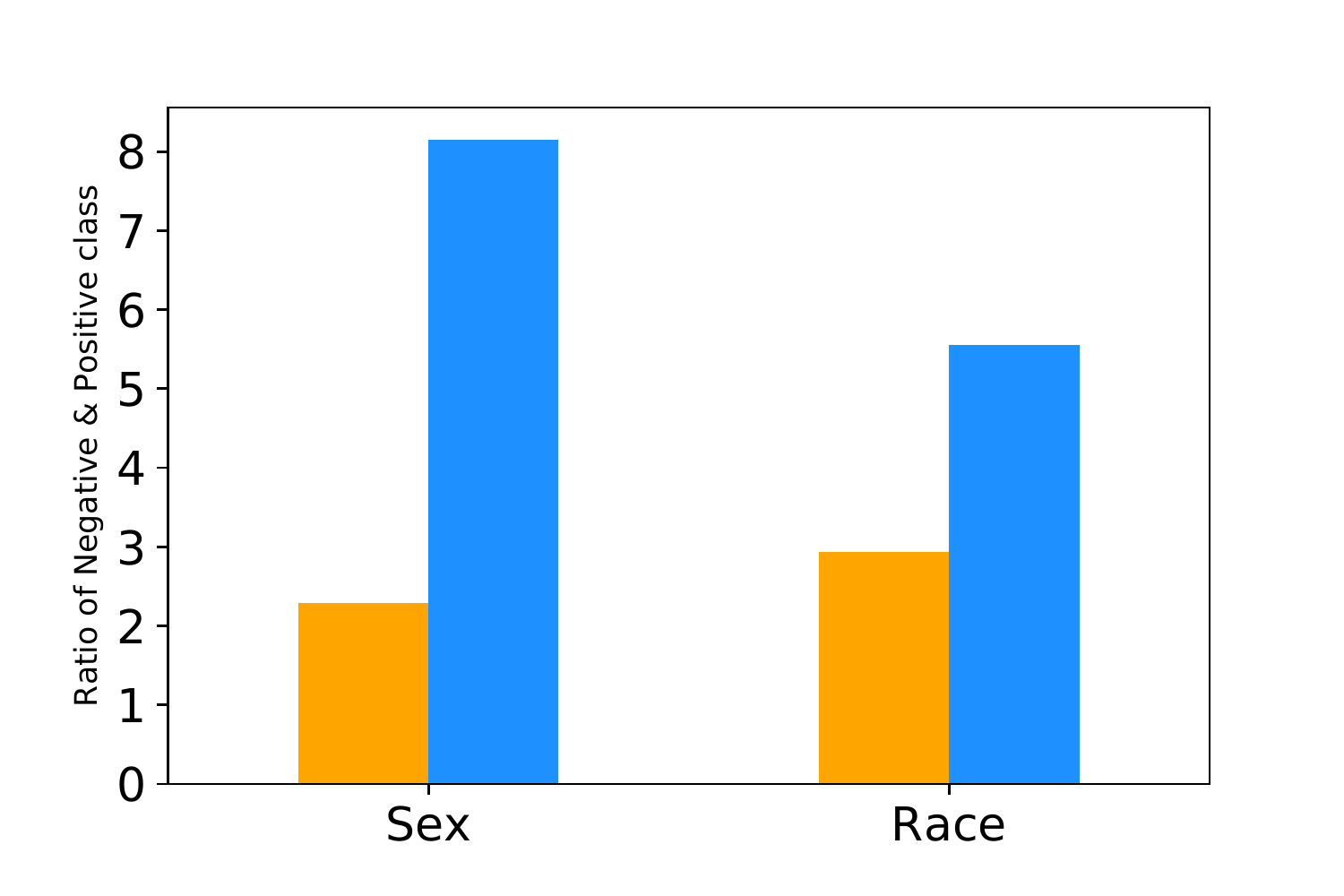}
\caption{ Ratio of negative and positive class for two protected attributes - sex and race for ``Adult'' dataset. ``Orange'' column is for Privileged group(Male,White) and ``Blue'' column is for unprivileged group(Female,Non-white).}
\label{Class_label}
\end{figure}

\bi
\item \textbf{Equal Opportunity Difference(EOD)}:  Difference of True Positive Rates(TPR) for unprivileged and privileged groups \cite{IBM}. 
\item \textbf{Average Odds Difference(AOD)}: Average of difference in False Positive Rates(FPR) and True Positive Rates(TPR) for unprivileged and privileged groups \cite{IBM}.
\ei

\begin{equation}
    TPR = TP/P = TP/(TP + FN)
\end{equation}

\begin{equation}
    FPR = FP/N = FP/(FP + TN)
\end{equation}

\begin{equation}
    EOD = TPR_U - TPR_P
\end{equation}

\begin{equation}
    AOD = [(FPR_U - FPR_P) + (TPR_U - TPR_P)] * 0.5
\end{equation}

EOD and AOD are computed using the input and output datasets to a classifier. A value of 0 implies that both groups have equal benefit, a value lesser than 0 implies higher benefit for the privileged group and a value greater than 0 implies higher
benefit for the unprivileged group. In this study, absolute value of these metrics have been considered.

\begin{table*}
\small
\caption{Description of the datasets used for the experiment.}
\label{dataset}
\begin{tabular}{@{}|ccccccc|@{}}
\toprule
\rowcolor[HTML]{C0C0C0} 
\textbf{Dataset}                                              & \textbf{\#Rows} & \textbf{\#Features} & \multicolumn{2}{c}{\cellcolor[HTML]{C0C0C0}\textbf{Protected Attribute}}                                                                    & \multicolumn{2}{c|}{\cellcolor[HTML]{C0C0C0}\textbf{Label}} \\
\rowcolor[HTML]{C0C0C0} 
\textbf{}                                                     & \textbf{}       & \textbf{}           & \textbf{Privileged}                                                 & \textbf{Unprivileged}                                                 & \textbf{Favorable}            & \textbf{Unfavorable}        \\ \midrule
\begin{tabular}[c]{@{}c@{}}Adult Census\\ Income\end{tabular} & 48,842          & 14                  & \begin{tabular}[c]{@{}c@{}}Sex-Male\\ Race-White\end{tabular}       & \begin{tabular}[c]{@{}c@{}}Sex-Female\\ Race-Non-white\end{tabular}   & High Income                   & Low Income                  \\ \midrule
Compas                                                        & 7,214           & 28                  & \begin{tabular}[c]{@{}c@{}}Sex-Female\\ Race-Caucasian\end{tabular} & \begin{tabular}[c]{@{}c@{}}Sex-Male\\ Race-Not Caucasian\end{tabular} & Did not reoffend              & Reoffended                  \\ \midrule
\begin{tabular}[c]{@{}c@{}}German Credit\\ Data\end{tabular}  & 1,000           & 20                  & Sex-Male                                                            & Sex-Female                                                            & Good Credit                   & Bad Credit                  \\ \midrule
Default Credit                                                & 30,000          & 23                  & Sex-Male                                                            & Sex-Female                                                            & Default Payment - Yes         & Default Payment - No        \\ \midrule
Heart Health                                                  & 297             & 14                  & Age-Young                                                           & Age-Old                                                               & Not Disease                   & Disease                     \\ \bottomrule
\end{tabular}
\end{table*}

Depending upon the notion of fairness, there are various fairness metrics also. The statistical notion of fairness in binary classification mainly comes from the confusion matrix - a table that is often used to describe the accuracy of a classification model. If there are two confusion matrices for two groups - privileged and unprivileged (see Table \ref{Confusion_Matrix}), all the fairness metrics try to find the difference of True Positive Rate and False Positive Rate for those two groups from those two matrices \cite{corbettdavies2017algorithmic,feldman2014certifying,hardt2016equality,alex2016fair,kleinberg2016inherent,IBM}. Beutel et al. commented that all of these fairness metrics suffer from three shortcomings \cite{beutel2019putting}-

\bi
\item These metrics ignore the class distribution for privileged and unprivileged groups. As a case study, Figure \ref{Class_label} shows the ratio of negative(low income) and positive(high income) class for two protected attributes - sex and race for ``Adult'' dataset. ``Orange'' column is for Privileged group(sex- male, race - white) and ``Blue'' column is for unprivileged group(sex- female, race - non-white). The figure shows the uneven distribution of positive and negative classes for unprivileged and privileged groups.

\item These metrics do not consider the sampling of the data. But incorrect sampling creates data imbalance which may lead to incorrect measurement of bias. 

\item These metrics ignore the cost of misclassification. For example, in case of credit card approval software, assigning bad credit score to an applicant who has actual good credit score is less costlier than assigning good credit score to an applicant who has actual bad credit score.

\ei

In this work, several steps are taken to overcome those shortcomings. Most of the prior works have either used AOD or EOD, we have used both of them for our study as we compared our approach with previous works \cite{IBM}. Instead of depending on only those two metrics, the concept of \textit{situation testing} was used to find discrimination \cite{Galhotra_2017}. In the context of binary classification, \textit{situation testing} is the process of verifying whether model prediction changes for same data point with changed protected attribute value \cite{10.5555/3060832.3061001}. While measuring the performance of Fairway, we used random sampling of data for ten times to overcome the sampling problem. Cost of misclassification is not solved because that is application specific and requires domain knowledge.

\section{Dataset Description}
\label{Datasets}

In this experiment, five datasets from UC Irvine Machine Learning Repository have been used. All the datasets are quite popular in fairness domain and used by previous SE researchers\cite{Galhotra_2017,Udeshi_2018,chakraborty2019software}. A brief description of the datasets are given -

\bi

\item Adult Census Income - This dataset contains records of 48,842 people. The class label is yearly income \cite{ADULT}. It is a binary classification dataset where the prediction task is to determine whether a person makes over 50K a year. There are fourteen attributes among them two are protected attributes. 


\item Compas - This is a dataset containing criminal history, demographics, jail and prison time,  and COMPAS (which stands for Correctional Offender Management Profiling for Alternative Sanctions) risk scores for defendants from Broward County \cite{COMPAS}. The dataset contains 7,214 rows and twenty-eight attributes. Among them there are two protected attributes. 


\item German Credit Data - This dataset contains records of 1,000 people and binary class labels (Good Credit or Bad Credit) \cite{GERMAN}. There are twenty attributes among them one is protected. 


\item Default Credit - There are 30,000 records of default payments of people from Taiwan \cite{DEFAULT}. Binary class label is Default Payment ``Yes'' or ``No''. There are twenty-three attributes among them one is protected. 


\item Heart Health -  The Heart Dataset from the UCI ML Repository contains fourteen features from 297 adults \cite{HEART}. The goal is to accurately predict whether or not an individual has
a heart condition. 


\ei

Table \ref{dataset} gives an overall description of all five datasets. These are binary classification datasets. Like most of the prior research\cite{10.1007/978-3-642-33486-3_3,NIPS2017_6988,hardt2016equality}, we used \textit{Logistic Regression} model on these datasets . But our approach is applicable for any classification model.

\section{The ``Fairway'' method}
\label{Methodology}

As stated above, the  Fairway algorithm is a combination of the pre-processing and in-processing approach to make machine learning software fairer.  

\subsection{Why not Remove the Protected Attributes?}

This section describes one of the methods
we explored, before arriving at Fairway.

When we think of prediction model discriminating over a protected attribute, the first solution which comes to mind is that why not train the model without that protected attribute. Being novice in fairness domain, we tried that for the five datasets. Two of the datasets have two protected attributes (Adult, Compas - Sex, Race) and other three datasets have only one protected attribute. We removed the protected attribute column from the train and test data so that the model has no information about that attribute. Surprisingly, there was almost no change in bias metrics even after that. 

Brun et al. have mentioned one reason behind this surprising result. They mentioned that if there is high correlation between attributes of the dataset, then even after removing the protected attribute, the bias stays \cite{FAIRWARE}. In 2016, Amazon created a model for same-day delivery service offered to Prime users around the major US cities\cite{Amazon}. But the model turned out to be highly discriminatory against black neighborhood. While training this model, ``Race'' attribute was not used but the model became biased against a certain ``Race'' because the ``Zipcode'' attribute highly correlates with ``Race''. The training data had ``Zipcode'' and the model induced ``Race'' from that. Initially, we also thought maybe correlation is the reason for our datasets also. But when we checked for the correlation between attributes, we found that bias is not coming from the correlation. 

For the datasets we are using here, the bias mainly comes from the class label. The data have been historically captured over the years. The classification was done by several human beings or algorithms - whether credit card gets approved or a person having a disease. Human bias or Algorithmic bias against certain sex or race reflected on predictions. In some cases, people of specific race or sex were unfairly treated. Thus the historical records have improper labels for some portion of data. 

This is to say that  even if we remove the ``protected attribute'' column, bias still remains. For removal of bias, we need to find out those data points having improper labels.

Finally, we can summarize different ways of a model acquiring bias from training data -

\bi

\item If in the training data, the class labels are related to any of the protected attributes, while training, a model can acquire that bias. If there is no protected attribute but other correlated attributes which affect the decision, then also model may become biased.

\item Kamishima et al. reported a reason for unfairness called ``Underestimation'' \cite{10.1007/978-3-642-33486-3_3}. It happens when  a trained model is not fully converged due to the finiteness of the size of the training data set. They defined a new metric called the underestimation index (UEI) based on the Hellinger distance to find ``Underestimation''. According to them, this occurs very rarely. So, we did not try to find UEI for our datasets.

\item Bias may come from unfair sampling of training data or unfair labeling of the training data. For the five datasets used in this study, the main reason of bias is unfair labeling of some data points. In this work, data has been randomly sampled ten times to make sure bias does not come from improper sampling.

\ei

 \begin{figure}[!t]
\centering
\includegraphics[width=.8\linewidth]{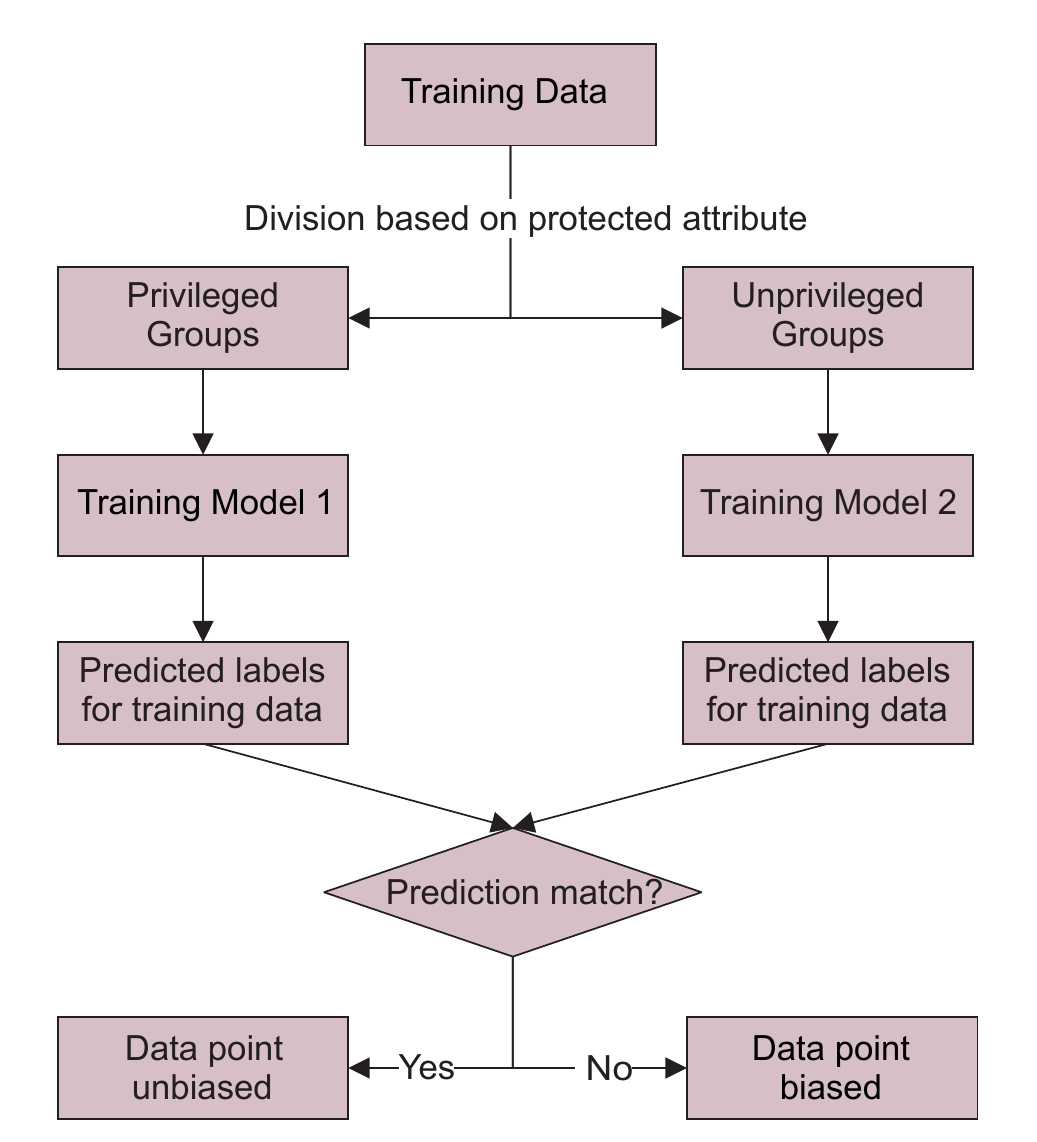}
\caption{ Pre-processing technique for bias removal from training data}
\label{Bias_removal}
\end{figure}

\subsection{Removal of Ambiguous (Biased) Data Points}
\label{Removal_of_biased_data_points}

Depending upon the protected attribute, there is a privileged group and an unprivileged group in each dataset. Which group is privileged and which group is unprivileged depend on the application. For example:
\bi
\item
In credit card applications, ``Male'' might be considered privileged and ``Female'' as unprivileged;
\item
In criminal prediction,
  ``White'' people might be considered privileged and ``non-white'' as unprivileged.
  \ei
  In this step, we try to find and remove the data points which are responsible for creating the bias based on the protected attribute. We call these data points the \textbf{ambiguous} data points. 

Fig. \ref{Bias_removal} describes the approach we applied to find out the ambiguous data points depending on the protected attribute. We divide the training data into two groups based on the protected attribute - privileged and unprivileged. Then we train two separate models on those two groups. Once we get the two trained models, for all the training data points, we check the prediction of these two models:
\bi
\item
If the prediction matches in both cases, the data point being examined is unbiased. 
\item
If two models contradict each other for a data point, there is a possibility of this data point being biased, this is an ambiguous data point. We remove that data point from training data. Later we will describe why this works and how to validate.
\ei
We call this data cleaning process as ``Bias Removal'' from training data. Once we are done removing the probable biased data points, we train a new model on the rest of the training data and make prediction using that model. Table \ref{pre_processing} shows the total number of rows in each dataset and the number of rows we removed. We see that at most we lose 15\% of training data after bias removal step. Later we will show that this does not affect much the performance of the prediction model.

We remove the ambiguous(bias causing) data points by constructing two separate logistic regression models conditioned upon the protected attribute of the dataset. Let's assume the original data points are denoted as X where $x_1,x_2,x_3,....,x_n$ are the attributes of the dataset and the protected attribute is denoted as s (s = $x_k$, where k is a number between 1 to n) and $\hat{y}$ is the model prediction. The original dataset is further divided into subsets based on the values of a protected attribute, in this case, $X_1 \subset X \forall s = 1$ and $X_2 \subset X \forall s = 0$. We use these two subsets to build two logistic regression models such as - 

\begin{equation}
    p(\hat{y} = 1 | s = 1) = \beta_0 + \beta_{1}x_1 + \beta_{2}x_2 + .... + \beta_{n-1}x_{n-1}
\end{equation}
\begin{equation}
   p(\hat{y} = 1 | s = 0) = \beta^{'}_0 + \beta^{'}_{1}x_1 + \beta^{'}_{2}x_2 + .... + \beta^{'}_{n-1}x_{n-1}
\end{equation}

\begin{equation}
    f_1(x) = \log_e{\frac{p(\hat{y} = 1 | s = 1)}{p(\hat{y} = 0 | s = 1)}}
\end{equation}
\begin{equation}
    f_2(x) = \log_e{\frac{p(\hat{y} = 1 | s = 0)}{p(\hat{y} = 0 | s = 0)}}
\end{equation}

Next, we use these logistic regression models to check for each training data point, by retaining the data points where 
\[\forall x \in X \left( f_1(x_1) == f_2(x_1)\right) \] This results in retaining only the data points where there is no contradiction about the models' outcome irrespective of data distribution conditioned upon the protected attribute, thus removing the data points which add ambiguity to the model and introduce bias into the model's prediction.

\begin{table}[!h]
\caption{ \#Rows = Total number of Rows, \#Dropped Rows = Total number of rows detected as ambiguous(biased)}
\label{pre_processing}
\small
\begin{tabular}{@{}|c|cccc|@{}}
\toprule
\rowcolor[HTML]{C0C0C0} 
{\color[HTML]{333333} \textbf{Dataset}}          & {\color[HTML]{333333} \textbf{\begin{tabular}[c]{@{}c@{}}Protected\\ Attribute\end{tabular}}} & {\color[HTML]{333333} \textbf{\#Rows}}                                  & {\color[HTML]{333333} \textbf{\begin{tabular}[c]{@{}c@{}}\#Dropped\\ Rows\end{tabular}}} & {\color[HTML]{333333} \textbf{\begin{tabular}[c]{@{}c@{}}\% of\\ Rows\\ Dropped\end{tabular}}} \\ \midrule
\rowcolor[HTML]{FFFFFF} 
\cellcolor[HTML]{FFFFFF}                         & Sex                                                                                           & \cellcolor[HTML]{FFFFFF}{\color[HTML]{333333} }                         & 6,178                                                                                    & 12.6                                                                                           \\
\rowcolor[HTML]{FFFFFF} 
\multirow{-2}{*}{\cellcolor[HTML]{FFFFFF}ADULT}  & Race                                                                                          & \multirow{-2}{*}{\cellcolor[HTML]{FFFFFF}{\color[HTML]{333333} 48,842}} & 2,315                                                                                    & 4.7                                                                                            \\ \midrule
\rowcolor[HTML]{FFFFFF} 
\cellcolor[HTML]{FFFFFF}                         & Sex                                                                                           & \cellcolor[HTML]{FFFFFF}                                                & 1,128                                                                                    & 15.6                                                                                           \\
\rowcolor[HTML]{FFFFFF} 
\multirow{-2}{*}{\cellcolor[HTML]{FFFFFF}COMPAS} & Race                                                                                          & \multirow{-2}{*}{\cellcolor[HTML]{FFFFFF}7,214}                         & 724                                                                                      & 10.0                                                                                           \\ \midrule
\rowcolor[HTML]{FFFFFF} 
DEFAULT CREDIT                                   & Sex                                                                                           & 30,000                                                                  & 505                                                                                      & 1.7                                                                                            \\ \midrule
\rowcolor[HTML]{FFFFFF} 
HEART HELTH                                      & Age                                                                                           & 297                                                                     & 32                                                                                       & 10.8                                                                                           \\ \midrule
\rowcolor[HTML]{FFFFFF} 
GERMAN                                           & Sex                                                                                           & 1,000                                                                   & 38                                                                                       & 3.8                                                                                            \\ \bottomrule
\end{tabular}
\end{table}

In the five datasets we used, due to the pre-processing step we do not lose much of training (see Table~\ref{pre_processing}). But in case of other datasets or real-world scenarios, if too many data points are found biased and model prediction gets damaged due to this loss, then we would suggest relabeling of data points instead of removal. In such relabeling, any majority voting technique like k-NN can be used. Biased data points will be assigned a new class label depending on k nearest neighbor data points. Such relabeling
comes with an extra cost (finding distance for all the data points), so we recommend it to use only if model prediction is affected due to the removal of biased data points. This study does not include that experiment, but this could be an interesting direction for future work.

\subsection{What if there are two protected attributes? }

Fig. \ref{Bias_removal} shows the approach we applied for one protected attribute. But in some cases, there are more than one protected attribute in a dataset. Like - Adult and Compas datasets (Sex and Race). If we have two protected attributes, we divide the training data based on those two attributes into four groups  (two privileged and two unprivileged groups). Then we apply the similar logic to find the biased data points. We train four different models on those four groups and check their predictions match or not. These models are not used for prediction, they are used to find biased data points only. In the two datasets, we did not lose more than 16\% of training data with this approach.

As to handling more than two protected attributes, we do not explore it here, for the following reason. With our data sets, such ternary (or more) protection divides the data into unmanageable small regions.
Future research in this area would require case studies with much larger data sets.


\subsection{Model Optimization}
IBM has created a GitHub repo to combine some promising prior works on fairness domain\cite{AIF360}. The results show that most of the prior methods damage the performance of the model while making it fair. So, prediction performance and fairness are competitive goals\cite{berk2017convex}.
When there is a trade-off between competing performance goals, multi-objective optimization is the way to explore the goal space. In our case, the goal of such optimizer would be to  make the model
as fair as possible while also not degrading other performance measures such as recall or false alarm.

To explore such multiobjective optimization,
we divided the dataset into three groups - Training (70\%), Validation (15\%) and Test (15\%)\cite{seventy_thirty}. During the pre-processing step, we removed biased data points from the training set. After that Logistic Regression model is trained on the training set with the standard default parameters\footnote{In Scikit-Learn, those details are  C=1.0, penalty=`l2', solver=`liblinear', max\_iter=100.}. Then we used the  FAIR\_FLASH algorithm (discussed below) to find out the best set of parameters to achieve optimal value of four metrics  (Higher Recall, Lower False Alarm, Lower AOD, and Lower EOD) on the validation set. Finally, the tuned model is applied  on the test set.

Nair et al. proposed FLASH \cite{8469102}, a novel optimizer, that utilizes sequential model-based optimization(SMBO). The concept of SMBO is very simple. It starts with ``What we already know about the problem'' and then decides ``what should we do next''. The first part is done by a machine learning model and the second part is done by an acquisition function. Initially, a few points are randomly selected and measured. These points along with their performance measurements are used to build a model. Then the model is used to predict the performance measurements of other unevaluated points. This process continues
until a stopping criterion is reached. FLASH improves over traditional SMBO as follows:

\bi

\item FLASH models each objective as a separate Classification and Regression Tree (CART) model. Nair et al. report that the CART
algorithm can scale much better than other model
constructors (e.g. Gaussian Process Models).

\item FLASH replaces the actual evaluation of all combinations of parameters(which can be a very slow process) with a \textit{surrogate evaluation}, where the CART decision trees are used to guess the objective scores (which is a very fast process). Such guesses may be inaccurate but, as shown by Nair et al., such guesses can rank guesses in (approximately) the same order as that generated by other, much slower, methods~\cite{nair2017using}.
\ei
FLASH was invented to solve software configuration problem and it performed faster than more traditional optimizers such as Differential Evolution\cite{Storn1997DifferentialE} or NSGA-II\cite{996017}. For our work, we modified  FLASH and generated  FAIR\_FLASH that seeks  best parameters for \textit{Logistic regression} model with four goals - higher recall, lower false alarm, lower AOD, lower EOD. Algorithm~\ref{FLASH_code} shows the pseudocode of FAIR\_FLASH. It has two layers - one learning layer and one optimization layer. When training data arrives, the estimator in the learning layer is being trained, and the optimizer in optimizing layer provides better parameters to the learner to help improve the performance of estimators. Such trained learner is evaluated on the validation data afterward. Once some stopping criteria is met, the generated learner is then passed to the test data for final testing.

\begin{algorithm}[!t]
\hspace{0.2cm}\begin{lstlisting}[xrightmargin=5.0ex,mathescape,frame=none]
def FAIR_FLASH():
    # pick a number of data into build_pool, evaluate the build_pool,
    # and put the rest into rest_pool
    while life > 0:
        # build CART model by using build_pool
        next_point = max(model.predict(rest_pool))
        build_pool += next_point
        rest_pool -= next_point
        if model.evaluate(next_point) < max (build_pool):
            life -= 1
    return max(build_pool)
    

\end{lstlisting}
\caption{Pseudocode of FAIR\_FLASH inspired from ~\cite{8469102}}
\label{FLASH_code}
\end{algorithm}

\begin{figure}
\centering
\includegraphics[width=.8\linewidth]{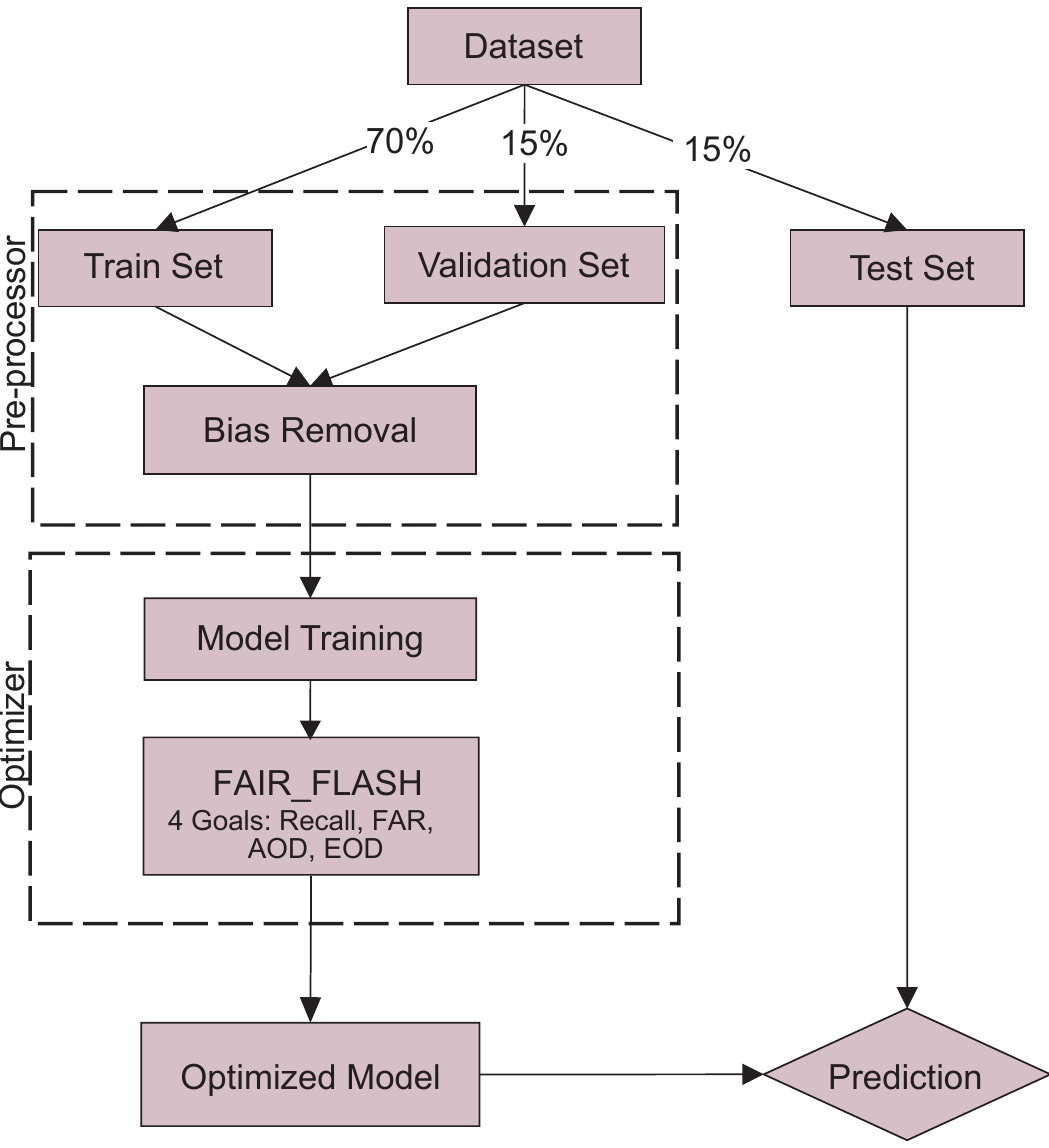}
\caption{ Block diagram of Fairway. For details on FAIR\_FLASH, see Algorithm~\ref{FLASH_code}. }
\label{FairWay}
\end{figure}

In summary, Fairway consists of two parts - bias removal from training data and model optimization to make trained model fair.  Fig. \ref{FairWay} shows an overview of the method.

\section{Results}
\label{Results}
Our results are structured around six research questions. For all the results, we repeated our experiments ten times with data shuffling and we report the median. 

\newenvironment{RQ}{\vspace{2mm}\begin{tcolorbox}[enhanced,width=3.3in,size=fbox,colback=blue!5,drop shadow southeast,sharp corners]}{\end{tcolorbox}}

\begin{figure*}
\includegraphics[width=\textwidth]{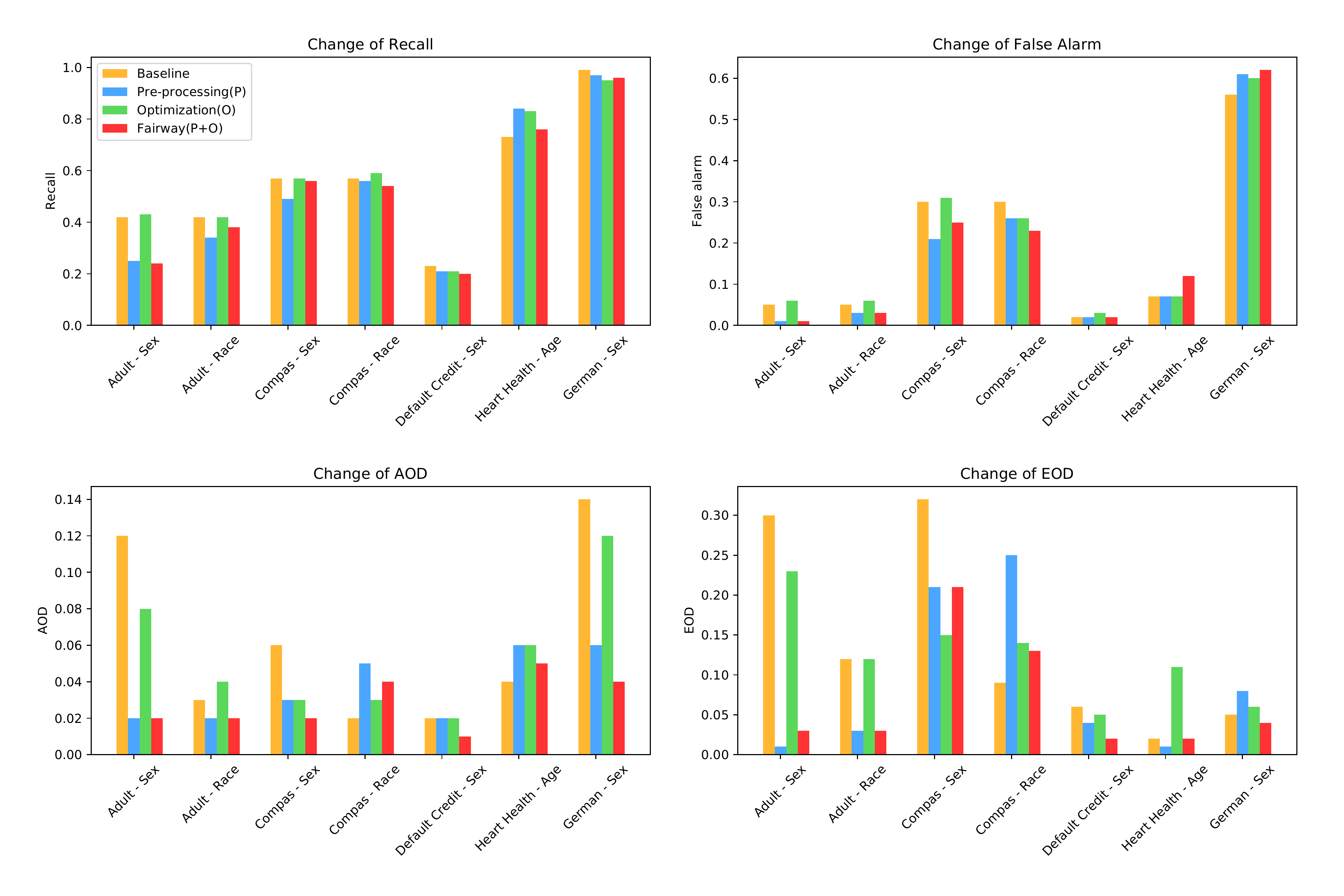}
\caption{Performance and fairness metrics for (a)~default state in Orange;  (b)~after pre-processing in Blue;
(c)~after just optimization in Green;
and (d)~after performing pre-processing + optimization in Red.  In these charts,
{\em higher} recalls are 
{\em better} while for all other scores, {\em lower values}
are {\em better}. }
\label{Three_methods}
\end{figure*}

\begin{RQ}
{\bf RQ1.} What is the problem with just using standard learners?
\end{RQ}
The premise of the paper is our methods offer some improvement over common practices. To justify that we first need to show that there are open issues with standard methods. We trained a logistic regression model with default scikit-learn parameters and tested on the five datasets. The ``Orange'' column in Fig. \ref{Three_methods} shows the results achieved using that model. The recall is higher the better, and false alarm, AOD, EOD are lower the better. Recall and False alarm are showing the prediction performance of the model. The high value of fairness metrics(AOD, EOD) in all five datasets signifies that model prediction is not fair means depending upon protected attribute, privileged group is getting advantage over unprivileged group. We treat this results as baseline for our experiment. We need to make the prediction fair without much damaging the performance. 

\begin{RQ}
{\bf RQ2.} How well does Pre-processing improve the results?
\end{RQ}
Fairway is a two-part procedure- data pre-processing (ambiguity removal) and learner optimization(FAIR\_FLASH). It is reasonable to verify the contribution of both parts.  Accordingly \textbf{RQ2} tests the effects of just doing ambiguity removal. 

Before training Logistic regression model, training data was cleaned to remove ambiguous data points(having improper labels) using the approach mentioned in section \ref{Removal_of_biased_data_points}. Table \ref{pre_processing} shows this step causes loss of maximum 15\%  of the training data. After that logistic regression model was trained on remaining data points and tested. The ``Blue'' column in Fig. \ref{Three_methods} shows the results achieved using that model. We see minor damage in recall for some cases and significant improvement in case of fairness metrics (lower AOD, EOD). It is evident that pre-processing the data before model training makes the model prediction fairer.

\begin{table*}
\caption{Comparison of Fairway with prior algorithms. Recall is higher the better. False alarm, AOD and EOD are lower the better. ``Gray'' cells show improvement and ``Black'' cells show damage. ``White'' cells show no change.}
\footnotesize
\begin{tabular}{@{}|cccccc
>{\columncolor[HTML]{343434}}c c
>{\columncolor[HTML]{C0C0C0}}c cc|@{}}
\toprule
\cellcolor[HTML]{C0C0C0}                                                                                            & \cellcolor[HTML]{C0C0C0}                                   & \cellcolor[HTML]{C0C0C0}                                               & \multicolumn{2}{c}{\cellcolor[HTML]{C0C0C0}\textbf{Recall}}                                   & \multicolumn{2}{c}{\cellcolor[HTML]{C0C0C0}\textbf{False alarm}}                 & \multicolumn{2}{c}{\cellcolor[HTML]{C0C0C0}\textbf{AOD}}                                      & \multicolumn{2}{c|}{\cellcolor[HTML]{C0C0C0}\textbf{EOD}}                        \\
\multirow{-2}{*}{\cellcolor[HTML]{C0C0C0}\textbf{Algorithm}}                                                        & \multirow{-2}{*}{\cellcolor[HTML]{C0C0C0}\textbf{Dataset}} & \multirow{-2}{*}{\cellcolor[HTML]{C0C0C0}\textbf{Protected Attribute}} & \cellcolor[HTML]{C0C0C0}\textbf{Before} & \cellcolor[HTML]{C0C0C0}\textbf{After}              & \cellcolor[HTML]{C0C0C0}\textbf{Before} & \cellcolor[HTML]{C0C0C0}\textbf{After} & \cellcolor[HTML]{C0C0C0}\textbf{Before} & \textbf{After}                                      & \cellcolor[HTML]{C0C0C0}\textbf{Before} & \cellcolor[HTML]{C0C0C0}\textbf{After} \\ \midrule
                                                                                                                    &                                                            & Sex                                                                    & 0.42                                    & \cellcolor[HTML]{343434}{\color[HTML]{FFFFFF} 0.38} & 0.05                                    & {\color[HTML]{FFFFFF} 0.07}            & 0.12                                    & 0.03                                                & 0.30                                    & \cellcolor[HTML]{C0C0C0}0.04           \\
                                                                                                                    & \multirow{-2}{*}{Adult}                                    & Race                                                                   & 0.42                                    & 0.42                                                & 0.05                                    & {\color[HTML]{FFFFFF} 0.08}            & 0.03                                    & \cellcolor[HTML]{FFFFFF}0.03                        & 0.12                                    & \cellcolor[HTML]{C0C0C0}0.09           \\
                                                                                                                    &                                                            & Sex                                                                    & 0.57                                    & \cellcolor[HTML]{343434}{\color[HTML]{FFFFFF} 0.56} & 0.30                                    & {\color[HTML]{FFFFFF} 0.32}            & \cellcolor[HTML]{FFFFFF}0.06            & 0.03                                                & 0.32                                    & \cellcolor[HTML]{C0C0C0}0.07           \\
                                                                                                                    & \multirow{-2}{*}{Compas}                                   & Race                                                                   & 0.57                                    & \cellcolor[HTML]{C0C0C0}0.59                        & 0.30                                    & {\color[HTML]{FFFFFF} 0.32}            & 0.02                                    & 0.01                                                & 0.09                                    & \cellcolor[HTML]{C0C0C0}0.03           \\
\multirow{-5}{*}{\textbf{Optimized Preprocessing\cite{NIPS2017_6988}}}                                                                  & German                                                     & Sex                                                                    & 0.99                                    & \cellcolor[HTML]{343434}{\color[HTML]{FFFFFF} 0.97} & 0.56                                    & {\color[HTML]{FFFFFF} 0.61}            & 0.14                                    & 0.12                                                & 0.04                                    & \cellcolor[HTML]{C0C0C0}0.03           \\ \midrule
                                                                                                                    &                                                            & Sex                                                                    & 0.42                                    & \cellcolor[HTML]{343434}{\color[HTML]{FFFFFF} 0.41} & 0.05                                    & {\color[HTML]{FFFFFF} 0.07}            & 0.12                                    & 0.02                                                & 0.30                                    & \cellcolor[HTML]{C0C0C0}0.03           \\
                                                                                                                    & \multirow{-2}{*}{Adult}                                    & Race                                                                   & 0.42                                    & \cellcolor[HTML]{343434}{\color[HTML]{FFFFFF} 0.40} & 0.05                                    & {\color[HTML]{FFFFFF} 0.08}            & 0.03                                    & 0.01                                                & 0.12                                    & \cellcolor[HTML]{C0C0C0}0.02           \\
                                                                                                                    &                                                            & Sex                                                                    & 0.57                                    & \cellcolor[HTML]{343434}{\color[HTML]{FFFFFF} 0.55} & 0.30                                    & {\color[HTML]{FFFFFF} 0.34}            & 0.06                                    & 0.03                                                & 0.32                                    & \cellcolor[HTML]{C0C0C0}0.12           \\
                                                                                                                    & \multirow{-2}{*}{Compas}                                   & Race                                                                   & 0.57                                    & 0.57                                                & 0.30                                    & {\color[HTML]{FFFFFF} 0.37}            & 0.02                                    & 0.01                                                & 0.09                                    & \cellcolor[HTML]{C0C0C0}0.03           \\
\multirow{-5}{*}{\textbf{Reweighing(Pre-processing)\cite{Kamiran2012}}}                                                               & German                                                     & Sex                                                                    & 0.99                                    & \cellcolor[HTML]{343434}{\color[HTML]{FFFFFF} 0.94} & 0.56                                    & {\color[HTML]{FFFFFF} 0.60}            & 0.14                                    & 0.10                                                & 0.04                                    & \cellcolor[HTML]{C0C0C0}0.03           \\ \midrule
                                                                                                                    &                                                            & Sex                                                                    & 0.42                                    & \cellcolor[HTML]{343434}{\color[HTML]{FFFFFF} 0.41} & 0.05                                    & \cellcolor[HTML]{C0C0C0}0.04           & 0.12                                    & 0.01                                                & 0.30                                    & \cellcolor[HTML]{C0C0C0}0.02           \\
                                                                                                                    & \multirow{-2}{*}{Adult}                                    & Race                                                                   & 0.42                                    & 0.42                                                & 0.05                                    & {\color[HTML]{FFFFFF} 0.07}            & 0.03                                    & 0.01                                                & 0.12                                    & \cellcolor[HTML]{C0C0C0}0.02           \\
                                                                                                                    &                                                            & Sex                                                                    & 0.57                                    & \cellcolor[HTML]{343434}{\color[HTML]{FFFFFF} 0.53} & 0.30                                    & {\color[HTML]{FFFFFF} 0.35}            & 0.06                                    & 0.04                                                & 0.32                                    & \cellcolor[HTML]{C0C0C0}0.06           \\
                                                                                                                    & \multirow{-2}{*}{Compas}                                   & Race                                                                   & 0.57                                    & \cellcolor[HTML]{343434}{\color[HTML]{FFFFFF} 0.52} & 0.30                                    & {\color[HTML]{FFFFFF} 0.36}            & 0.02                                    & 0.02                                                & 0.09                                    & \cellcolor[HTML]{C0C0C0}0.06           \\
\multirow{-5}{*}{\textbf{\begin{tabular}[c]{@{}c@{}}Adversial Debiasing\cite{zhang2018mitigating}\\ (In-processing)\end{tabular}}}            & German                                                     & Sex                                                                    & 0.99                                    & \cellcolor[HTML]{343434}{\color[HTML]{FFFFFF} 0.94} & 0.56                                    & {\color[HTML]{FFFFFF} 0.60}            & 0.14                                    & 0.12                                                & 0.04                                    & \cellcolor[HTML]{FFFFFF}0.04           \\ \midrule
                                                                                                                    &                                                            & Sex                                                                    & 0.42                                    & \cellcolor[HTML]{343434}{\color[HTML]{FFFFFF} 0.21} & 0.05                                    & \cellcolor[HTML]{C0C0C0}0.02           & 0.12                                    & 0.03                                                & 0.30                                    & \cellcolor[HTML]{C0C0C0}0.04           \\
                                                                                                                    & \multirow{-2}{*}{Adult}                                    & Race                                                                   & 0.42                                    & \cellcolor[HTML]{343434}{\color[HTML]{FFFFFF} 0.23} & 0.05                                    & \cellcolor[HTML]{C0C0C0}0.02           & 0.03                                    & 0.02                                                & 0.12                                    & \cellcolor[HTML]{C0C0C0}0.10           \\
                                                                                                                    &                                                            & Sex                                                                    & 0.57                                    & \cellcolor[HTML]{C0C0C0}0.62                        & 0.30                                    & {\color[HTML]{FFFFFF} 0.34}            & 0.06                                    & 0.01                                                & 0.32                                    & \cellcolor[HTML]{C0C0C0}0.03           \\
                                                                                                                    & \multirow{-2}{*}{Compas}                                   & Race                                                                   & 0.57                                    & \cellcolor[HTML]{C0C0C0}0.61                        & 0.30                                    & {\color[HTML]{FFFFFF} 0.34}            & 0.02                                    & \cellcolor[HTML]{FFFFFF}0.02                        & 0.09                                    & \cellcolor[HTML]{C0C0C0}0.07           \\
\multirow{-5}{*}{\textbf{\begin{tabular}[c]{@{}c@{}}Reject Option Classification\cite{Kamiran:2018:ERO:3165328.3165686}\\ (Post-processing)\end{tabular}}} & German                                                     & Sex                                                                    & 0.99                                    & \cellcolor[HTML]{343434}{\color[HTML]{FFFFFF} 0.94} & 0.56                                    & {\color[HTML]{FFFFFF} 0.61}            & 0.14                                    & 0.04                                                & 0.04                                    & \cellcolor[HTML]{C0C0C0}0.01           \\ \midrule
                                                                                                                    &                                                            & Sex                                                                    & 0.42                                    & \cellcolor[HTML]{343434}{\color[HTML]{FFFFFF} 0.24} & 0.05                                    & \cellcolor[HTML]{C0C0C0}0.01           & 0.12                                    & 0.02                                                & 0.30                                    & \cellcolor[HTML]{C0C0C0}0.03           \\
                                                                                                                    & \multirow{-2}{*}{Adult}                                    & Race                                                                   & 0.42                                    & \cellcolor[HTML]{343434}{\color[HTML]{FFFFFF} 0.38} & 0.05                                    & \cellcolor[HTML]{C0C0C0}0.03           & 0.03                                    & 0.02                                                & 0.12                                    & \cellcolor[HTML]{C0C0C0}0.03           \\
                                                                                                                    &                                                            & Sex                                                                    & 0.57                                    & 0.57                                                & 0.30                                    & \cellcolor[HTML]{C0C0C0}0.25           & 0.06                                    & 0.02                                                & 0.32                                    & \cellcolor[HTML]{C0C0C0}0.21           \\
                                                                                                                    & \multirow{-2}{*}{Compas}                                   & Race                                                                   & 0.57                                    & \cellcolor[HTML]{343434}{\color[HTML]{FFFFFF} 0.54} & 0.30                                    & \cellcolor[HTML]{C0C0C0}0.23           & 0.02                                    & \cellcolor[HTML]{343434}{\color[HTML]{FFFFFF} 0.04} & 0.09                                    & \cellcolor[HTML]{C0C0C0}0.13           \\
\multirow{-5}{*}{\textbf{\begin{tabular}[c]{@{}c@{}}Fairway\\ (Pre-processing + In-processing)\end{tabular}}}       & German                                                     & Sex                                                                    & 0.99                                    & \cellcolor[HTML]{343434}{\color[HTML]{FFFFFF} 0.96} & 0.56                                    & {\color[HTML]{FFFFFF} 0.62}            & 0.14                                    & 0.04                                                & 0.04                                    & 0.04                                   \\ \bottomrule
\end{tabular}
\label{Algo_comparison}
\end{table*}

\begin{RQ}
{\bf RQ3.} How well does Optimization improve the results?
\end{RQ}
Moving on from \textbf{RQ2}, the third research question is to check the effect of just optimization(no pre-processing). 

To do that, we tuned the Logistic regression model parameters using FAIR\_FLASH to optimize the model for higher recall, lower false alarm and lower fairness metrics(AOD, EOD). Then the tuned model was used for prediction. The ``Green'' column in Fig. \ref{Three_methods} shows the results achieved using that model. We see that in cases of prediction performance(recall, false alarm) it performs similar or better than pre-processing but in case of fairness metrics(AOD, EOD), pre-processing does better. So, optimized learner is significantly better than baseline learner but combining pre-processing may perform even better. 

\begin{RQ}
{\bf RQ4.} How well does Fairway improve the results?
\end{RQ}

Our fourth research question explores the effect of Fairway which is a combination of pre-processing and optimization. 

The ``Red'' column in Fig. \ref{Three_methods} shows the results achieved after applying Fairway. Fairway is performing  better than pre-processing and optimization in most of the cases. 
For example:
\bi
\item In case of Adult dataset, for the protected attribute race, Fairway achieves almost similar recall with optimization but much better in the other three metrics.
\item In case of Default Credit dataset, for the protected attribute sex, Fairway is providing best results for all four metrics. 
\ei

In some cases, recall is slightly damaged. But overall, Fairway is making the model fair without much affecting the performance. So, pre-processing the data before model training and tuning the model while training both are important.

\begin{RQ}
{\bf RQ5.} How well does Fairway perform compared to previous fairness algorithms?
\end{RQ}

We have decided to compare our approach Fairway with some popular previous algorithms described in section \ref{Removing_Bias}. We chose five such algorithms (all from IBM AIF360) which we thought could be representative of the works done before. Table \ref{Algo_comparison} shows the results for three datasets - Adult, Compas, and German. It shows the change of recall, false alarm and two fairness metrics AOD, EOD before and after the algorithms are applied. In most of the cases, Fairway is performing better or the same with prior algorithms in case of reducing ethical bias (AOD,EOD). In case of false alarm, Fairway has less number of black cells showing damage. Like Fairway, previous algorithms also slightly damage the recall metric. In some situations, this may become a matter of concern. We see a scope of improvement here where future researchers should focus.

 We have performed scott-knott significance test and A12 effect size test for comparison. For AOD, Fairway performs better in 2/5 cases and for EOD, in 3/5 cases. Here better means result is statistically significantly better. For the rest of the cases, although having the same rank, improvement is between 10\%-25\%.  Also, Fairway wins on false alarm for all cases and keeps the same recall in  3/5 cases and damages in 2/5 . And when Fairway loses in recall, it does not lose by much (10\%-12\%).

Fairway is not just another bias mitigation approach. It differs from prior works in several ways -

\bi

\item The first part of Fairway is finding bias in training data. So, even before model training, Fairway shows which data points in the training data have improper/biased labels and can affect prediction in future. If labeling was done by human reviewers, it leads to finding bias in human decisions. Instead of blindly trusting the ground truth of training data, Fairway can be used to find bias in the ground truth.

\item Prior bias mitigation algorithms come from the core concepts of machine learning. Software practitioners having little ML knowledge may face difficulties to use these algorithms\cite{Holstein_2019}. In case of Fairway, users can clearly see how two different models trained on privileged and unprivileged groups give different predictions on biased data points. This makes Fairway much comprehensible. FAIR\_FLASH gives user the flexibility to choose which parameters to optimize. In this paper, Logistic regression model is used. But FAIR\_FLASH is easily extensible for other classification models. So, FAIR\_FLASH is adjustable too.

\item Fairway is a combination of bias testing and mitigation. This is described in \textbf{RQ6}. 

\ei


\begin{RQ}
{\bf RQ6.} Can Fairway be used as a combined tool for detection and mitigation of bias?
\end{RQ}

\begin{figure}[!t]
\begin{center}
\includegraphics[width=\linewidth]{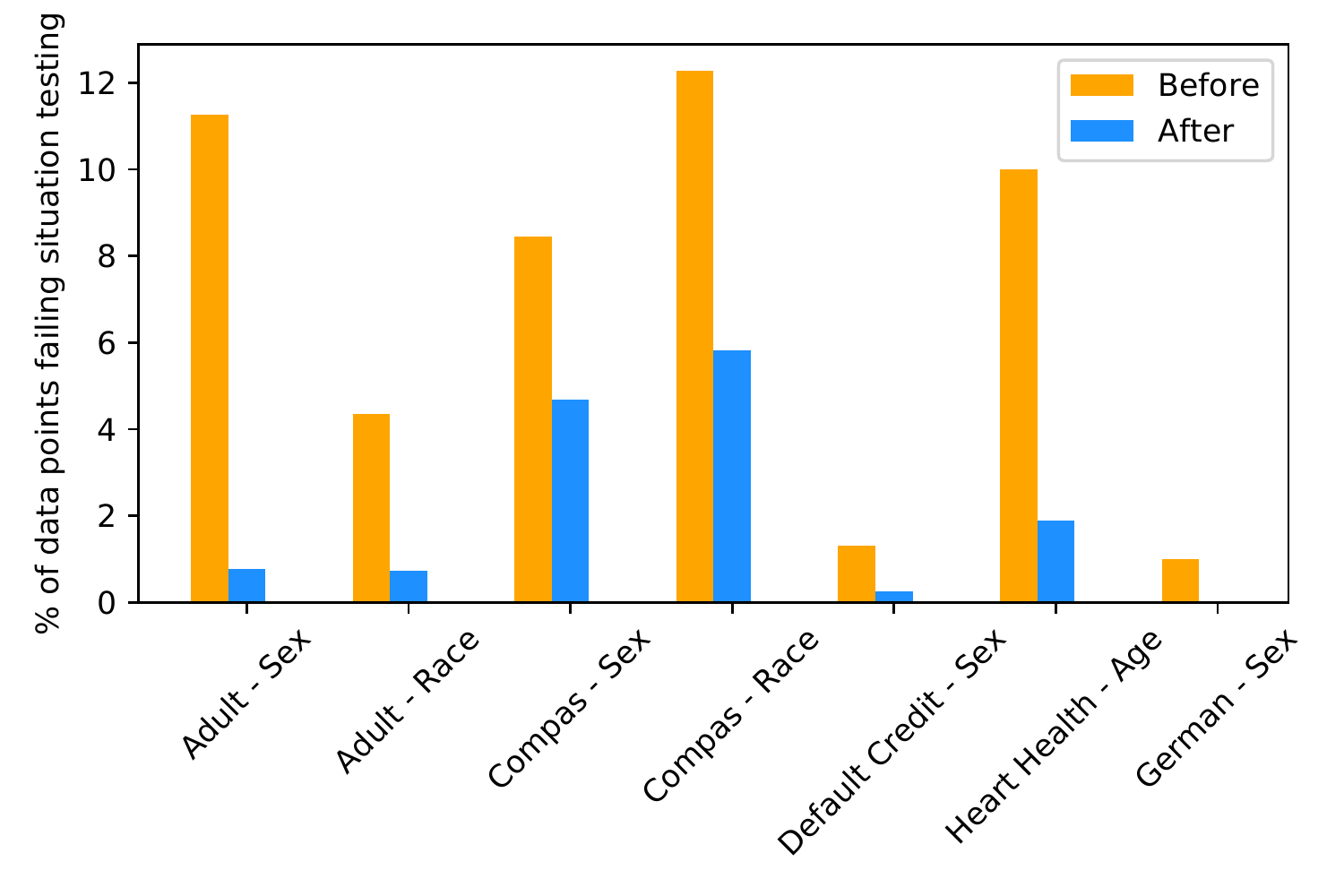}
\end{center}
\caption{ Percentage change of data points failing situation testing (showing bias) before and after pre-processing.  }
\label{FairWay_Verify}
\end{figure}

In section \ref{Background And Previous Work}, it is shown that there are mainly two types of previous works done by researchers - finding the bias in AI software and mitigating the bias. As per our knowledge, we are the first one to combine these two. Fairway finds the data points which have unfair labeling in the training data and remove those data points so that prediction is not affected by protected attribute. We used \textit{Situation testing} \cite{10.5555/3060832.3061001} to verify whether after bias removal, the role of a protected attribute on the prediction changes or not. we switched the protected attribute value for all the remaining data points  (e.g. we changed
Male to Female and Female to Male).  Then we checked whether these changes lead to prediction changes or not. If the prediction changes for a data point, we say that it fails situation testing. Figure \ref{FairWay_Verify} shows the percentage of data points failing situation testing before and after pre-processing step of Fairway:

\bi
\item The ``orange” and ``blue'' columns show results before/after  applying Fairway.
\item In all cases, the values on the blue  column are far smaller than orange column.
\ei

So, Fairway can find the data points responsible for bias in the training data. Now, it is an engineering decision to set the threshold of what percentage of training data can be ambiguous where prediction may change depending on the protected attribute value. Fairway provides the percentage and depending on the application, user can decide whether bias is present in the system or not. So, Fairway can be applied as a discrimination finder tool. If discrimination is above the tolerable threshold, then Fairway can be applied for removing bias from training data and optimizing model without damaging predictive performance. So, Fairway can be used as a combined tool for detection and mitigation of discrimination or ethical bias. One unique feature of Fairway is it is \textbf{model-agnostic}. It finds bias by verifying prediction of a model and mitigates bias by cleaning training data and tuning model parameters. So, it can work for any black box model. As Fairway only works on the output space of a model,  it can be easily used in industrial purposes where revealing core algorithm of the underlying model is not possible.

So, to summarize the results, we say that we have explained the reasons of bias in the five datasets we used. We have developed a comprehensible method \textit{Fairway} which can remove bias from training data and the model. Unlike prior works, \textit{Fairway} is not just a bias mitigation approach, it is a combined tool for ground truth validation, bias detection and mitigation.

\section{Threats to Validity}
\label{Threats_to_Validity}

\bi
\item \textbf{Sampling Bias} - We have used five datasets from UCI machine learning repository where most of prior works in fairness domain use only one or two datasets. These are well-known datasets and used by previous researchers in ML and software fairness domain. It is an open issue if these data sets reflect an interesting range
of fairness issues for other data sets. In future work, we would explore more data sets. 

\item \textbf{Evaluation Bias} - We have used two fairness metrics - EOD and AOD. We have mentioned the drawbacks of fairness metrics which only consider the TPR and FPR and neglect the class distribution. Recent work has deduced a new fairness metric called \textit{Conditional Equality of Opportunity} to overcome this drawback \cite{beutel2019putting}. Conditional Equality of Opportunity is defined for conditioning on every feature and finding the opportunity gap for privileged and unprivileged groups.   In future work, we would explore more performance criteria.

\item \textbf{Construct Validity} - In our work we trained different models on privileged and unprivileged groups. The datasets contained one or two protected attributes, so our method is feasible. All the prior works we have seen treated each protected attribute individually.  We have shown how to deal with two protected attributes. In future work, we would explore larger data sets with more protected attributes.

\item \textbf{External Validity} - Fairway is limited to classification models which are very common in AI software. We are currently working on extending it to Regression models. 
In future work, we would extend this work to other kinds of data mining problems; e.g. to text mining or video processing systems.
\ei

\section{Conclusion}
\label{Conclusion}

 We have explained how a model acquires bias from improper labels of training data and have demonstrated an approach called ``Fairway'' which removes ``ethical bias'' from the training data and optimizes a trained model for fairness and performance. We have shown that Fairway is comprehensible and can be used as a combined tool for detection and mitigation of bias. Unlike some prior ML works, Fairway is not just a bias mitigation tool, it validates ground truth labels, finds bias and mitigates bias. We have made the source code of ``Fairway'' publicly available for software researchers and practitioners. To the best of our knowledge, we claim this is the first work in SE domain which concentrates on mitigating ethical bias from software and making software fair using optimization methods augmented with some data pre-processing.  In future, we hope more and more software researchers will work on this domain and industries will consider publishing more datasets. When that data becomes available, it would be appropriate to rerun this study.
\cite{NIPS2017_7062}

\balance
\bibliographystyle{IEEEtran} 
\bibliography{main}

\end{document}